\documentclass{aa}
\usepackage{graphicx}
\begin{document}
\title{Constraints on the halo density profile using HI flaring in the outer Galaxy}
\author{C.A. Narayan, K. Saha, \& C.J. Jog}
\offprints{C. A. Narayan}
\institute{Department of Physics, Indian Institute of Science, Bangalore 560012, India.\\
\email{chaitra@physics.iisc.ernet.in}}
\date{Received; accepted}

\abstract{
The observed flaring of HI disk in the outer region of galaxies has been 
used in the past to determine the shape of the dark matter halo. Previous 
studies based on this concept suggest a slightly oblate halo 
(axis ratio $\sim$ 0.8) for our Galaxy. We reinvestigate this problem by 
calculating the HI scaleheight in the outer Galaxy to a larger
radial distance, and by studying its dependence on the shape and the density 
profile of the halo. We 
find that a simple isothermal infinite halo of any shape- oblate or prolate, 
is not able to account for the observed flaring. Instead we 
show that a spherical halo with density falling faster than isothermal halo
 in the outer region provides a better fit to the observed HI flaring as well
 as the observed rotation curve of our Galaxy. 
 These halos have about 95$\%$ of their mass within a few hundreds of kpc.
For R$_{\circ}$ = 8.5 kpc and $\Theta_{\circ}$ = 220 kms$^{-1}$, the 
central density and core radius can be constrained to the 
range $\rho_{\circ}$ = 0.035 - 0.06 M$_{\odot}$ pc$^{-3}$ and R$_c$ = 8 - 10 kpc. Our claim for such 'finite-sized' spherical halos is supported by recent literature on numerical simulation studies of halo formation as well as analyses of SDSS data. 
}
\authorrunning{Narayan et al.}
\titlerunning{Constraints on the Halo density profile using HI flaring}
\maketitle

\section{Introduction}
 
It is a well-known observational fact that the atomic hydrogen layer flares in the outer Galaxy (Kulkarni, Heiles \& Blitz 1982; Knapp 1987; Wouterloot et al. 1990; Diplas \& Savage 1991; Merrifield 1992; Nakanishi \& Sofue 2003). HI flaring is also noticed in many external galaxies seen edge-on (Brinks \& Burton 1984; Olling 1996a; Matthews \& Wood 2003). A possible cause for flaring could be that the total gravitational force acting perpendicular to the disk plane decreases with radius while the velocity dispersion of HI is observed to be nearly constant (Lewis 1984). The contribution to the total perpendicular gravitational force comes mainly from the stellar disk, gas and the dark matter. While the stellar disk dominates within the optical region of a normal disk galaxy, the outer region is dominated by the dark matter. The disk and halo seem to dominate in different regions of a galaxy because of their different density distributions. Although both decrease from the center, the disk density decreases rapidly whereas the halo density decreases much more slowly, so that the halo extends to several times the size of the optical disk. Thus the halo plays a major role in determining the vertical disk structure beyond the optical region. This makes the outer Galactic HI layer sensitive to the mass and the density profile of the halo, and hence it can be used as a diagnostic to study the halo properties.

The first attempt towards studying the halo parameters using HI layer was made by Olling (1995), who has developed a model where the observed thickness of the HI layer can be used to predict the shape of the halo. This method has been used to show that the halos of NGC 4244 (Olling 1996b) and NGC 891 (Becqueart \& Combes 1997) are highly flattened with their axis ratios in the range of 0.2 - 0.4. Olling \& Merrifield (2000) use the same method to find an axis ratio of 0.8 for the halo of our Galaxy. However, on spanning a larger parameter space and considering more factors which affect the HI scaleheight, they find the halo to be closer to spherical for any R$_{\circ} >$ 7 kpc where  R$_{\circ}$ is the distance between sun and the Galactic center (Olling \& Merrifield 2001). Another measurement of the shape the halo of our Galaxy comes from an entirely different method. Using the kinematics of the tidal streams of stars surrounding the Galaxy, called the Sagittarius stream, Ibata et al. (2001) show that the Galactic dark halo is almost spherical. They conclusively rule out axis ratios $<$ 0.7 for our Galaxy. In general, there is a lot of interest now in determining the shape of the dark matter halo in a galaxy (see e.g., Natarajan 2002, and the references therein). 

The focus of this paper is to determine the halo parameters which provide the best fit (in the least square sense) to the observed HI scaleheight. The halo parameters we intend to explore are its central mass density $\rho_{\circ}$, core radius R$_c$, the shape (or axis ratio) $q$ and the power law index of the density profile $p$. We calculate the HI scaleheight, numerically, based on our self-consistent model for the Galaxy (Narayan \& Jog 2002) and compare the results with the observations. The gravitational force due to the dark matter halo is incorporated in the model as an external force acting on the disk. The effect of the disk gravity on the halo is thus neglected in this model so that the halo is taken to be rigid or non-responsive. The HI scaleheight is first obtained using the simplest profile for the halo - a spherical screened isothermal density profile. Subsequently, the shape of this halo is varied. Next we study the effects of including a non-isothermal halo by varying the power law index of its density profile. We find that the best fit to the data is obtained for a spherical halo whose density falls more rapidly ($p$=1.5-2) than that of an isothermal halo ($p$=1).
 
The formulation of the problem is described in Section~2 while the method 
of calculation and parameters used in the model are discussed in Section~3. 
Section 4 is devoted to the results of this paper, and in Sections 5 and 6, 
we present discussion and conclusions.

\section{Formulation of the problem} 
\subsection{Vertical equilibrium}

We treat the Galactic disk to consist of three major components: the stars, the interstellar atomic hydrogen gas HI, and the interstellar molecular hydrogen gas H$_2$, which are coupled gravitationally. We assume that the three components are axisymmetric and are in hydrostatic equilibrium in the $z$ direction. We use the galactic cylindrical co-ordinates $R, \phi, z$. Then the disk dynamics under the force field due to the rigid halo can be described by the Poisson equation, and the force equation along the normal to the plane (the equation of pressure equilibrium) for each component. The Poisson equation for the axisymmetric galactic system in cylindrical geometry is given as:

$$ \frac{\partial^2 \Phi_{tot}}{\partial z^2} + \frac{1}{R} 
\frac{\partial}{\partial R} (R \frac{\partial \Phi_{tot}}{\partial R}) \:
  = \:  4 \pi G \: (\sum_{i=1}^3 \rho_i  + \rho_h )
\eqno(1) $$

\noindent where $\rho_i$ with $i$ = 1 to 3  denotes the mass density for each disk component, $\rho_h$ denotes the same for the halo, and $\Phi_{total}$ denotes the potential due to the disk and the halo. For a disk with a flat rotation curve, the radial term in the above equation is identically equal to zero at the mid-plane ($z$ = 0). The rotation curve of our Galaxy is not observed to be strictly flat (Merrifield 1992; Honma \& Sofue 1997). But quantitatively, we find that the radial term contributes to less than 1 $\%$ change in the HI scaleheight. Hence, this term can be neglected and equation (1)  reduces to:

$$ \frac{\partial^2 \Phi_{tot}}{\partial z^2}  \: = \:
    4 \pi G \: (\sum_{i=1}^3 \rho_i + \rho_h )  \eqno(2) $$

The equation for pressure equilibrium in the vertical direction for each component is given by (e.g. Rohlfs 1977):

$$ \frac{\partial}{\partial z}(\rho_i \overline{v^2_z}_i) + 
 \rho_i \frac{\partial \Phi_{tot}}{\partial z} \: = \: 0 \eqno(3) $$

\noindent On combining the above two equations, we get the equation for the vertical equilibrium of each component in the disk under the field of the halo to be:

$$ \frac{\partial}{\partial z} \left [ \frac {\overline{(v^2_z)}_i} {\rho_i} 
\frac{\partial \rho_i}{\partial z} \right]
 = -4\pi G \: ( \sum_{i=1}^3  \rho_i + \rho_h )  \eqno(4) $$

\noindent Here, we assume that the vertical velocity dispersion is independent of distance from the disk plane, that is, the disk is taken to be isothermal. Note that the above treatment {\it does not} assume the disk to be thin. Under the thin disk approximation, the disk contribution to the radial term in equation (1) would drop out and equation (4) would reduce to :

$$\frac{\partial}{\partial z} \left [ \frac {\overline{(v^2_z)}_i} {\rho_i} 
\frac{\partial \rho_i}{\partial z} \right]  = -4\pi G \: ( \sum_{i=1}^3  \rho_i ) + \left . \frac{\partial K_z}{\partial z} \right |_{halo} \eqno(5) $$ 

\noindent where K$_z$ is the vertical force due to the halo. The above equation was used to calculate the stellar and gas scaleheights in the inner region of the Galaxy in our earlier paper (Narayan \& Jog 2002). We find that treating the HI disk to be thin in the present case would overestimate the HI scaleheight by as much as $10\%$. Hence, we have used the general approach for a thick disk for HI gas (eq.[4]) in the present paper.

\subsection{Dark matter halo}

We assume a four-parameter halo model as described by the following density profile (de Zeeuw \& Pfenniger 1988, Becquaert \& Combes 1997) 

$$ \rho(R,z) \: = \: \frac{\rho_{\circ}(q)} {\left(1 + \frac{m^2}{R^2_c(q)}\right)^p} \eqno(6) $$ 

\noindent where $\rho_{\circ}$ is the central mass density of the halo, $R_c$(q) is the core radius, $q$ is the axis ratio and $p$ is the index. Here by definition, $ m^2 = R^2 + {z^2}/{q^2} $ represents the surfaces of concentric ellipsoids. Note that $q$ = 1 would give rise to a spherical halo, while $q = c/a <1$ gives an oblate halo and $q = c/a >1$ describes a prolate halo, where $a$ is the axis in the disk plane and $c$ is that along the vertical direction.

By varying $p$ one can generate different halo density profiles.
$p$=1 gives a screened isothermal halo. The mass density here
goes as r$^{-2}$ at large radii (r$\gg$R$_c$). This leads to the
mass within a spheroid $M(r) \propto r$ and a flat rotation
curve. For $p$=1.5, $\rho \propto r^{-3}$ at large r just like
the NFW halo (Navarro et al. 1996) and this gives $M(r) \propto
log(r)$. i.e., the total mass goes over to infinity (similar to
the case of $p$=1) but more gradually. This family of halos
gives rise to a falling rotation curve for r$\gg$R$_c$. $p$=2
halos have their mass density falling much faster ($\propto
r^{-4}$ at large r). $M(r)$ saturates to a finite value, unlike
the other two cases. 
The rotation curve falls faster than the former class. 

\subsubsection{Halo shape : isothermal halo}

In the above equation $p$ = 1 gives a screened isothermal halo. Varying $q$ in this profile would give an axisymmetric 'pseudo-isothermal' halo of a different shape. When the shape of a halo of fixed mass changes to either prolate or oblate, its central density, $\rho_{\circ}$, and the core radius, R$_c$, are bound to change in order to conserve the mass. Therefore to know the exact density profile for any ellipsoid, we need to first calculate the $\rho_{\circ}$ and $R_c$ as functions of $q$.

 We find that $\rho_{\circ}(q)$ and $R_c(q)$ are related to their spherical counterparts by the following relations:

$$ \rho_{\circ}(q) = \rho_{\circ}(1) \frac{1}{q}\left ( \frac{e}{sin^{-1}e} \right)^3  $$ 
$$ R_c(q) = R_c(1) \left ( \frac{sin^{-1}e}{e} \right) \eqno (7)  $$    

\noindent These are  obtained by imposing the following two constraints : the mass within a thin spheroidal shell (Binney \& Tremaine 1987, pg 54) and the terminal velocity of the halo should be independent of $q$. Here $e$ is the eccentricity, and $e=\sqrt{1 - q^2}$ for the oblate case ($q<1$) ; and $e=\sqrt{1 - (1/q^2)}$ for the prolate case ($q > 1$). Figures (1a) and (1b) show the plots for $\rho_o(q)$ and $R_c(q)$ respectively as a function of oblateness and figures (1c) and (1d) give the corresponding plots for prolate shapes.
 
\subsection{Comparison with previous work}

We note that relations for $\rho_o(q)$ and $R_c(q)$ for the oblate case have been previously obtained by Olling (1995) (see figure 2 of his paper) but by using a different method. 
The rotation velocity at core radius and the terminal velocity are the two constraints used to derive the relations for $\rho_o(q)$ and $R_c(q)$ in his work. This method also yields similar results but the method is cumbersome and the relations are approximations.
 
In this respect, we find that our method is advantageous in the following ways : first, it gives simple and accurate relations for $R_c(q)$ and $\rho_o(q)$, and second, these relations can be used for the oblate as well for the prolate cases along with the appropriate relations between $e$ and $q$ as given above, whereas Olling considered only oblate halos.

The 'global approach' originally proposed by Olling (1995) and subsequently used by Olling \& Merrifield (2000, 2001), to calculate the HI scaleheight is a general one and can be used even if the stellar disk is truncated before the HI layer ends. The method used in this paper is the so-called 'local approach' (see Sect. 2.1), where the density obtained as a solution is based on the local graviational potential. This method, which was previously used by Spitzer (1942) and Bahcall (1984) to get classic results for the vertical disk distribution, is adopted in our work too because of its simplicity. The drawback of this method is that it does not yield self-consistent results (vertical density distribution and scaleheight of HI in this case) for truncated stellar disks. In our work (as in the works of Spitzer and Bahcall), the assumed density distributions of all the disk components and the halo are continuous. Hence the results are very close approximations of the exact self-consistent solutions (obtained by using the global approach) and are therefore valid.

\section{Calculations and input parameters}

Equation (4) represents the three coupled equations for the three disk components (stars, HI, and H$_2$) which are to be solved for the corresponding density distributions. The vertical density distribution for each component responding to the total potential of the disk and the halo, is solved for numerically as an initial value problem, using the fourth order Runge-Kutta method of integration (Press et al 1994). The details of this procedure are presented in our earlier paper (Narayan \& Jog 2002). At any radius R, the HWHM of the vertical density distribution is defined as the scaleheight. Repetition of the calculation at regular intervals of the radius  gives us the 'model' scaleheight curve.

The input parameters for the model for each disk component are its surface density and vertical velocity dispersion. The  surface densities for HI and H$_2$ are taken from the observations of Wouterloot et al. (1990). The stellar disk surface density is assumed to fall exponentially with distance from the center. The stellar disk mass and the surface density at any radius can be calculated using the following measured/inferred quantities : the stellar surface density at solar region $\Sigma_{\odot}$, the disk scalelength $R_d$ and the distance of sun from the center R$_{\circ}$. We use $\Sigma_{\odot}$ = 45 M$_{\odot} pc^{-2}$ which is consistent with 48 $\pm$ 9 M$_{\odot} pc^{-2}$ obtained by Kuijken \& Gilmore (1991) and 52 $\pm$ 13 M$_{\odot} pc^{-2}$ obtained by Flynn \& Fuchs (1994) for the total surface density, after the gas density is subtracted. We use the IAU recommended value for R$_{\circ}$ (=8.5 kpc) and $R_d$ is equal to 3.2 kpc (Mera et al. 1998) in accordance with the recent determinations of smaller disk scalelength for our Galaxy.

The stellar vertical dispersion is derived from the observations of radial dispersion by Lewis \& Freeman (1989) and the assumption that the ratio of the vertical to radial velocity dispersion is equal to 0.45 at all radii in the Galaxy, equal to its observed value in the solar neighbourhood as obtained from the analysis of the Hipparcos data (Dehnen \& Binney 1998, Mignard 2000). The vertical velocity dispersion for H$_2$ is taken to be 5 km s$^{-1}$ (Stark 1984, Clemens 1985). 

\subsection{HI velocity dispersion}

The HI velocity dispersion is observed to be almost constant with radius and is about 9$\pm$1 km s$^{-1}$ (Spitzer 1978; Malhotra 1995) in the inner Galaxy (upto solar circle). Beyond the solar circle, however, the dispersion is not yet measured.  A study of 200 external galaxies (Lewis 1984) shows that the observed dispersion has a very narrow range, about 8$\pm$1 km s$^{-1}$, consistent with observations of our Galaxy. Sicking's (1997) work shows that in two external galaxies, dispersion decreases slowly upto the outer edge of the HI layer. In a number of other galaxies, the velocity dispersion decreases and then saturates to a constant value of 7$\pm$1 km s$^{-1}$ (Shostak \& van der Kruit 1984; Dickey 1996; Kamphuis 1993). This decrease in dispersion is perhaps due to the lesser number density of supernovae in the outer region (McKee \& Ostriker 1977) . A major part of the observed dispersion is non-thermal in origin and the supernovae could be the major source for this whereas the thermal contribution comes upto just about 1 km s$^{-1}$ (Spitzer 1978).

In this work, the dispersion is taken to be 9 km s$^{-1}$ at 9 kpc, consistent with Malhotra (1995). Between 9-20 kpc the dispersion is allowed to decrease linearly with a slope of -0.2 km s$^{-1}$kpc$^{-1}$, as it is found to give the best fit to the data. Beyond 20 kpc, which is about twice the optical disk size, it is kept fixed at 7 km s$^{-1}$, equivalent to that observed in external galaxies. This is perhaps justified in the absence of direct observations in the outer Galaxy. 
 
\section{Results}

In this section we calculate the HI scaleheight versus radius in the outer Galaxy using the above parameters for the disk components and different density profiles for the dark matter halo. We then compare our results with observed HI scaleheight data. Of the various observations of HI scaleheight in the outer Galaxy by different authors, we find that results of Knapp (1987), Wouterloot et al (1990) and Merrifield (1992) are consistent with each other. Of these, Wouterloot et al give scaleheight values upto a very large distance (nearly 3R$_{\circ}$) with closest sampling (bin size of about 250 pc). Therefore we use their data to fit with the results from our model. Unfortunately, the error bars associated with these data points are not given. So, we compute least square (which is equivalent to $\chi^2$ with unit error bars) of the model-generated curve in order to measure its goodness of fit. Merrifield (1992) has pointed out that Wouterloot's data has to be corrected for beam-size effects but we note that the correction on the derived HI scaleheight is so small (fig 4 of Merrifield 1992) that it can be ignored in our study.

The first choice of dark matter profile used here is that proposed as a 
part of the complete mass model of the Galaxy by Mera et al. (1998) based 
on microlensing observations. It is a simple screened isothermal spherical 
halo ($p$ = 1) with $R_c$ = 5 kpc, the central density 
$\rho_0$= 0.035 M$_{\odot} pc^{-3}$ and the terminal rotation velocity = 220 km $s^{-1}$. Their stellar disk parameters are $\Sigma_{\odot}$ = 45 M$_{\odot} pc^{-2}$, $R_d$ = 3.2 kpc and R$_{\circ}$ = 8.5 kpc. Figure 2 (solid line) shows that the HI scaleheight obtained using  this profile matches well with observations upto about 20 kpc. But, beyond this region the calculated values fall much below the observed points, thereby giving a poor fit (a high value of least square).

\subsection{Shape of the halo}

Next, we consider whether halo of any different axisymmetric shape can improve the fit with the data. In doing so the remaining halo parameters and the disk parameters are kept unchanged. Such halos, especially those of oblate shape have gained popularity as discussed  in Section 1.  

Figure 2 shows the resulting scaleheight curves from our model for a range of values for $q$, the axis ratio. In addition to the oblate halos that are extensively studied in the literature, we also consider halos that are prolate shaped. This makes the study of the effect of halo shape on the HI flaring complete. Also, there is some evidence in recent literature in favour of prolate-shaped halos. For example, Ideta et al. (2000) find that prolate halo helps sustain warps better. Figure 2 shows that the scaleheight upto 16 kpc radius remains almost unaffected by the change in halo shape and the effect of shape is prominent only beyond 20 kpc. Also note that both the oblate and  the prolate cases reduce to that of the spherical halo under limiting conditions (i.e., for q = 1). A definite trend is observed as the shape is varied from oblate to prolate.  The oblate halo tends to reduce the scaleheight and this reduction increases with flattening. The effect is exactly opposite for the prolate-shaped halo. This is because the mid-plane halo density increases on compressing the halo (oblate shape) and decreases on elongating it (prolate shape). This results in a higher constraining force due to the oblate shape, and vice versa for the prolate shape.

It is clearly seen that the results from the pseudo-isothermal models (p=1;q$\neq$1) do not fit the observations well irrespective of the choice of shape and the axis ratio. The least square for the  oblate case increases with increasing oblateness. Though the curves generated by prolate halos also do not give a good fit, the least square is lower than that for the oblate case. The overall trend shown by the prolate curves matches with observations better but they are still far from giving a good fit to the data.

\subsection{Halo density profiles}

Since the variation in the shape of the halo does not lead to a good agreement between the model and the observations, we consider change in the density profiles as characterised by the index $p$ (eq.[6]). As $p$ is varied between 1, 1.5 and 2, the shape of the halos is kept spherical for the sake of simplicity. For each value of $p$, a realistic range of $\rho_o$ and $R_c$ is chosen to form a grid of ($\rho_o$,$R_c$) pairs. The core density $\rho_o$ is varied between 0.001-0.1 $M_{\odot}pc^{-3}$ in steps of 0.002 $M_{\odot}pc^{-3}$ and $R_c$ is varied between 4-15 kpc in steps of 100 pc. The total (disk+halo) circular speed at the solar point is calculated corresponding to each of these grid points. The HI scaleheight is calculated for only those grid points which give circular speed in the range determined by the relation between Galactic constants, $\Theta_{\circ}$ = (27$\pm$2.5)$R_{\circ}$ km s$^{-1}$ (Kerr \& Lynden-Bell 1986, Reid et al. 1999). This is the first constraint used to narrow down the number of possibilities for the best fit halo model. The second constraint is that the least square value of the model scaleheight curve should be minimum. We find that just a small subset ($\sim$10) of the entire set of grid points ($\sim$5000) gives minimum values of the least square. The final constraint imposed on this subset is that the total disk+halo rotation curves generated by these halos should show the main trends seen in the observed rotation curve (Merrifield 1992; Honma \& Sofue 1997) such as the rise beyond the solar point and fall beyond 2$R_{\circ}$. The above procedure is repeated for other values of index $p$.

For the specific choice of  $R_{\circ}$ = 8.5 kpc ($R_d$ = 3.2 kpc; 
$\Sigma_{\odot}$ = 45$M_{\odot}pc^{-2}$), the $\Theta_{\circ}$ is expected 
to be in the range 230$\pm$21 km s$^{-1}$. The least square value is 
found to be minimum for $p$ = 2 and $\rho_o$, $R_c$ are in the
range of 0.035-0.093 $M_{\odot}pc^{-3}$ and 7-9.5 kpc
respectively. These halos span the entire range of allowed
$\Theta_{\circ}$ ie, between 210-250 km s$^{-1}$. Figure 3 shows
the case of $\rho_o$ = 0.035 $M_{\odot}pc^{-3}$  and $R_c$ = 9.4
kpc. As the central density increases (and $R_c$ decreases), the
$\Theta_{\circ}$ increases progressively. In each case,  the
total rotation curve rises from the solar point, reaches a peak
value and then falls beyond. But the rise becomes more gradual,
the peak shifts towards centre and the fall becomes steeper as
the central density increases (for the higher end of
$\Theta_{\circ}$), thus growing more and more inconsistent with
the observed rotation curve. Therefore we find that halos with
$\rho_o$ = 0.035-0.06 $M_{\odot}pc^{-3}$ and $R_c$ = 8-9.5 kpc ,
which give $\Theta_{\circ}$ = 210 to 230 km s$^{-1}$ give a good
 fit to the HI data as well as produce realistic rotation curves.
  
The choice of $p$=1.5 for the halo profile also gives a reasonable fit to the data points, though not as good as for $p$=2. The fall in the rotation curve generated by $p$=1.5 is so gradual that it almost appears to be constant in the region of interest. This is however certainly within the error bars of the observed rotation curve (Honma \& Sofue 1997). The least square values for the $p$=1 halos are so large (for physically meaningful $\rho_o$, $R_c$) that they do not pass the very first constraint used to obtain the best fit halos.

Thus our model shows a preference for $1.5 < p \leq 2$ in the density profile for the halo (eq.[6]), corresponding to a density fall-off proportional to r$^{-\beta}$ with $3 <\beta \leq 4$, at very large radii ($R\gg R_c$). For $p$=2, the density falls to $10^{-4}$ of the central density at just 10 core radii ($\sim$100kpc). In comparison, the fractional
fall-off for isothermal case ($p=1$) is only $10^{-2}$ at the same distance. This steeper density fall can give rise to the rapid observed flaring between 20-24 kpc. See Fig. 4a for a comparison between $p$=1 and $p$=2 density profiles.
Note that at large radii ($R\gg R_c$), $p$=1.5 goes over to the popular NFW profile for the dark halo.

The p=2 halo gives a finite mass on integrating upto infinity.
For example, the best fit halo has a total mass of 2.8$\times
10^{11}$ M$_{\odot}$ of which about 90\% lies within 100kpc. In
contrast, the mass of an isothermal halo is linearly
proportional to the distance of integration and therefore
becomes infinite at infinity (see Fig. 4b). Since most of the
mass of p=2 halos are confined to within a few hundreds of kpc
(or a few decades of core radii), 
these can be regarded as naturally truncated halos or 'finite sized' halos. 
This concept has very important consequences in the cosmological scenario.

In Figure 4c, we plot the rotation curves corresponding to $p=1$
and $2$ cases. This illustrates the difference between these
two model cases especially at large radii, and is in a form that
is directly amenable to future observational checks.
In the case of $p=1$, the rotation curve becomes flat at large radii in keeping
with the linearly rising mass of the halo (compare with Fig. 4b); 
whereas in the case of $p=2$, the rotation curve begins to drop 
beyond $\sim$ 12 kpc.

\subsection{Comparison with other deductions}
Observationally, the vast database of SDSS has allowed researchers to check the large-scale behaviour of the dark matter halos. Prada et al. (2003) find that the radial density 
profile $\propto r^{-3}$ as predicted by most modern cosmological models, 
is consistent with the observed velocity dispersion of satellites 
of isolated galaxies. This corresponds to $p$=1.5 in our model. 
Fischer et al. (2000) and McKay et al. (2002), on the other hand find 
that the halo mass density falls off as r$^{-4}$ at very large radii 
(corresponding to $p$=2 in our model) i.e., for r $\gg$ 260 h$^{-1}$ kpc, 
which is the minimum size limit for an isolated galaxy. 
This limit corresponds to about 400kpc within which 95\% of 
mass is confined for our best fit halos. These support the overall 
trend in the results from our work. However, we need to be careful in 
comparing our results with the above SDSS studies which are on isolated 
galaxies while our Galaxy has at least one massive close neighbour.

The faster than isothermal fall-off is also supported by many numerical simulation studies in the literature. In their simulations on formation of dark halos, Avila-Reese et al. (1999) find that most halos tend to have density profiles $\propto r^{-\beta}$ where $\beta$ falls in the range 2.5 - 3.8, in the outer region. Bournaud et al. (2003) find from their simulations, that dark halos should extend to at least ten times further than their stellar disks, in order to be able to explain the formation of tidal dwarf galaxies. 
We note that our best fit halo 
certainly extends much beyond the corresponding limit of $\sim$ 120kpc.
 Further support for the choice of $p=2$ comes from  studies which show that the halo core radius is comparable to the optical/Holmberg radius for a galaxy (Salucci 2001). As $p$ increases, the best fit value for R$_c$ also increases. Hence $p$=2 halos give the largest value for R$_c$ (8-10 kpc) which is closest to the Homberg radius estimated for our Galaxy (e.g., Binney \& Merrifield 1998).

In the present work, the halos that provide a reasonable fit 
to the observed flaring have $p$ 
in the range of 1.5-2 but were all spherical in shape. This result is 
consistent with recent findings on the shape of dark halo of our Galaxy 
(Ibata et al. 2001; Olling \& Merrifield 2001). These recent studies 
indicate that the shape is nearly spherical (see Sect. 1). Thus 
we find that the results for both density profile and shape ($p$ and $q$) 
for the dark halo of our Galaxy as deduced from the HI flaring, 
are consistent with recent observational evidence and theoretical works. 

\section{Discussion}

\noindent {\bf (1) Asymmetry in the Galaxy :}
A very crucial assumption in our model is that the Galactic disk and the 
halo are axisymmetric. This is done for simplicity, as was also done by 
previous authors (Olling \& Merrifield 2000, 2001). 
We note, however, that the outer Galaxy shows asymmetry or lopsidedness.
There is observational evidence for kinematical lopsidedness
where the cut-off in the fourth and the first quadrants differs
by 25 km s$^{-1}$ (Burton 1988), and also for spatial
lopsidedness where the measurable column density extends much
farther out in the north than in the south (upto 4 R$_{\odot}$
and 2.2 R$_{\odot}$ respectively)- see Merrifield (2002), also
see Nakanishi \& Sofue (2003). Thus the analysis and the
conclusions from our paper are largely based on the northern
data and that is the limitation in the analysis of the present paper.

\noindent {\bf (2) Galactic constants :}
We have built the stellar disk model based on the IAU-recommended values 
for the galactic constants - R$_{\circ}$ = 8.5 kpc and 
$\Theta_{\circ}$ = 220 km s$^{-1}$. It would definitely be interesting 
and also worthwhile to know how the results for the halo density profile 
would vary with the assumed galactic constants. 
For example, Olling \& Merrifield (2001) find the effect of varying 
these constants on the inferred axis ratio of the halo. Unfortunately, 
all the observational inputs for our model, like the HI and H$_2$ surface 
densities, HI scaleheight and the stellar velocity dispersion, are based 
on the IAU-recommended galactic constants and rescaling them for 
different values of the constants is beyond the scope of this paper.

\noindent {\bf (3) Self-gravity of the gas :} 
It is interesting to check the change brought about by excluding the 
HI self-gravity on its vertical scaleheight. The difference is not 
negligible, it is seen to be about 10-20$\%$ within the optical disk 
(R$<$ 4R$_d$) and also beyond R $>$ 6R$_d$. This could be because these 
regions are dominated by the stellar disk and by the dark matter 
respectively. For the region where 4 $<$ R/R$_d$ $<$ 6, the difference 
is substantial ($\sim$50$\%$) suggesting that in this range, the gas gravity 
is crucial in negotiating the hydrostatic equilibrium for the HI layer. Thus 
neglecting it may lead to a serious overestimate of the HI scaleheight 
in general at all radii in the outer Galaxy 
(as already cautioned by Olling 1995) and particularly in the intermediate 
range of radii.

\section{Conclusions}

We calculate the HI scaleheight in the outer Galaxy using a Galactic disk model taking the dark matter halo also into account. In the outer Galaxy, the dark matter halo is the key component that decides the scaleheight of HI, hence we calculate the radial variation in the HI scaleheight as a function of the shape and density profile of the halo. Based on the method of least squares we show that neither oblate nor prolate-shaped isothermal halos can provide a good fit to the observations. 

Instead, the best agreement with data is provided by a spherical halo and a density profile that is $\propto r^{-\beta}$ with $3<\beta \leq4$ in the peripheral parts of the Galaxy ($1.5<p \leq 2$). In such a halo, the density falls off stepper than the-often-used isothermal halo. The rotation curves produced by these best-fit halos are also in good agreement with the observed one. This result seems to be in good agreement with the recent trend seen in the literature on the numerical simulations of halo formation, as well as the halo density profiles deduced from the SDSS data.

\noindent {\bf Acknowledgements}  

We would like to thank the referee, Albert Bosma, for his critical 
comments and the many suggestions,  which have vastly improved the 
quality of this paper. We also thank Anish Roshi 
and Rekhesh Mohan for discussions on HI observations and their analysis.
K.S. would like to thank the CSIR-UGC, India for a Senior research fellowship.
 
\newpage

\noindent {\bf {References}}

\noindent Avila-Reese, V., Firmani, C., Klypin, A., \& Kravtsov, A.V. 1999, MNRAS, 310, 527

\noindent Bahcall, J.N. 1984, ApJ, 276, 156

\noindent Becquaert, J.-F., Combes, F. 1997, A\&A, 325, 41

\noindent Binney, J., \& Merrifield, M. 1998, Galactic Astronomy (Princeton: Princeton Univ. Press)

\noindent Binney, J., \& Tremaine, S. 1987, Galactic Dynamics (Princeton: Princeton Univ. Press)

\noindent Bournaud, F., Duc, P.-A., \& Masset, F. 2003, A\&A, 411, L469

\noindent Burton, W.B. 1988, in Galactic and Extragalactic Radio
Astronomy, eds. G.L. Verschuur \& K.I. Kellermann (New York: Springer-Verlag)

\noindent Brinks, E., \& Burton, W.B. 1984, A\&A, 141, 195

\noindent Clemens, D.P. 1985, ApJ, 295, 422

\noindent Dehnen, W., \& Binney, J. 1998, MNRAS, 298, 387

\noindent Dickey, J.M. 1996, in I.A.U. Symp. 169, Unsolved problems of the Milky Way, eds. L. Blitz \& P. Teuben (Dordrecht: Kluwer), 489

\noindent Diplas, A., \& Savage, B.D. 1991, ApJ, 377, 126

\noindent Fischer, P. et al. 2000, AJ, 120, 1198

\noindent Flynn, C., \& Fuchs, B. 1994, MNRAS, 270, 471

\noindent Honma, M., \& Sofue, Y. 1997, PASJ, 49, 453

\noindent Ibata, R. et al. 2001, ApJ, 551, 294

\noindent Ideta, M., Hozumi, S., Tsuchiya, T., \& Takizawa, M. 2000, MNRAS, 311, 733 

\noindent Kamphuis, J.J. 1993, PhD Thesis, University of Groningen

\noindent Kerr, F., \& Lynden-Bell, D. 1986, MNRAS, 221, 1023

\noindent Knapp, G.R. 1987, PASP, 99, 1134

\noindent Kuijken, K., \& Gilmore, G. 1991, ApJ, 367, L9

\noindent Kulkarni, S.R., Heiles, C., \& Blitz, L. 1982, 259, L63  

\noindent Lewis, B.M. 1984, ApJ, 285, 453

\noindent Lewis, J.R., \& Freeman, K.C. 1989, AJ, 97, 139

\noindent Malhotra, S. 1995, ApJ, 448, 138

\noindent Matthews, L.D., \& Wood, K. 2003, ApJ, 593, 721

\noindent McKee, C.F., \& Ostriker, J.P. 1977, ApJ, 218, 148

\noindent McKay, T.A. et al. 2002, ApJ, 571, L85 

\noindent Mera, D., Chabrier, G., \& Schaeffer, R. 1998, A\&A, 330, 953

\noindent Merrifield, M.R. 1992, AJ, 103, 1552

\noindent Mignard, F. 2000, A \& A, 354, 522

\noindent Nakanishi, H., \& Sofue, Y. 2003, PASJ, 55, 191

\noindent Narayan, C.A., \& Jog, C.J. 2002, A\&A, 394, 89 

\noindent Natarajan, P. 2002, The shape of galaxies and their
dark halos (Yale cosmology workshop), (Singapore: World Scientific)

\noindent Navarro, J.F., Frenk, C.S., \& White, S.D.M. 1996, ApJ, 462, 563

\noindent Olling, R.P. 1995, AJ, 110, 591

\noindent Olling, R.P. 1996a, AJ, 112, 481

\noindent Olling, R.P. 1996b, AJ, 112, 457

\noindent Olling, R.P., \& Merrifield, M.R. 2000, MNRAS, 311, 361

\noindent Olling, R.P., \& Merrifield, M.R. 2001, MNRAS, 326, 164

\noindent Prada, F., et al. 2003, ApJ, 598, 260

\noindent Press, W.H., Flannery, B.P., Teukolsky, S.A., \& Vetterling, 
W.T. 1986, Numerical Recipes (Cambridge: Cambridge Univ. Press), chap. 6.

\noindent Reid, M.J., Readhead, A.C.S., Vermeulen, R.C., \& Treuhaft R.N. 1999, ApJ, 524, 816

\noindent Rohlfs, K. 1977, Lectures on density wave theory
 (Berlin: Springer-Verlag)

\noindent Salucci, P. 2001, MNRAS, 320, L1

\noindent Shostak, G.S., \& van der Kruit, P.C. 1984, A\&A, 132, 20

\noindent Sicking, F.J. 1997, PhD Thesis, University of Groningen

\noindent Spitzer, L. 1942, ApJ, 95, 329

\noindent Spitzer, L. 1978, Physical Processes in the Interstellar Medium
 (New York: John Wiley)

\noindent Stark, A.A. 1984, ApJ, 281, 624

\noindent Wouterloot, J.G.A., Brand, J., Burton, W.B., \& Kwee, K.K. 1990,
A\&A, 230, 21

\noindent de Zeeuw, T. \& Pfenniger, D. 1988, MNRAS, 235, 949

\bigskip

\newpage

\bigskip
\begin{figure}
{\rotatebox{270}{\resizebox{9cm}{9cm}{\includegraphics{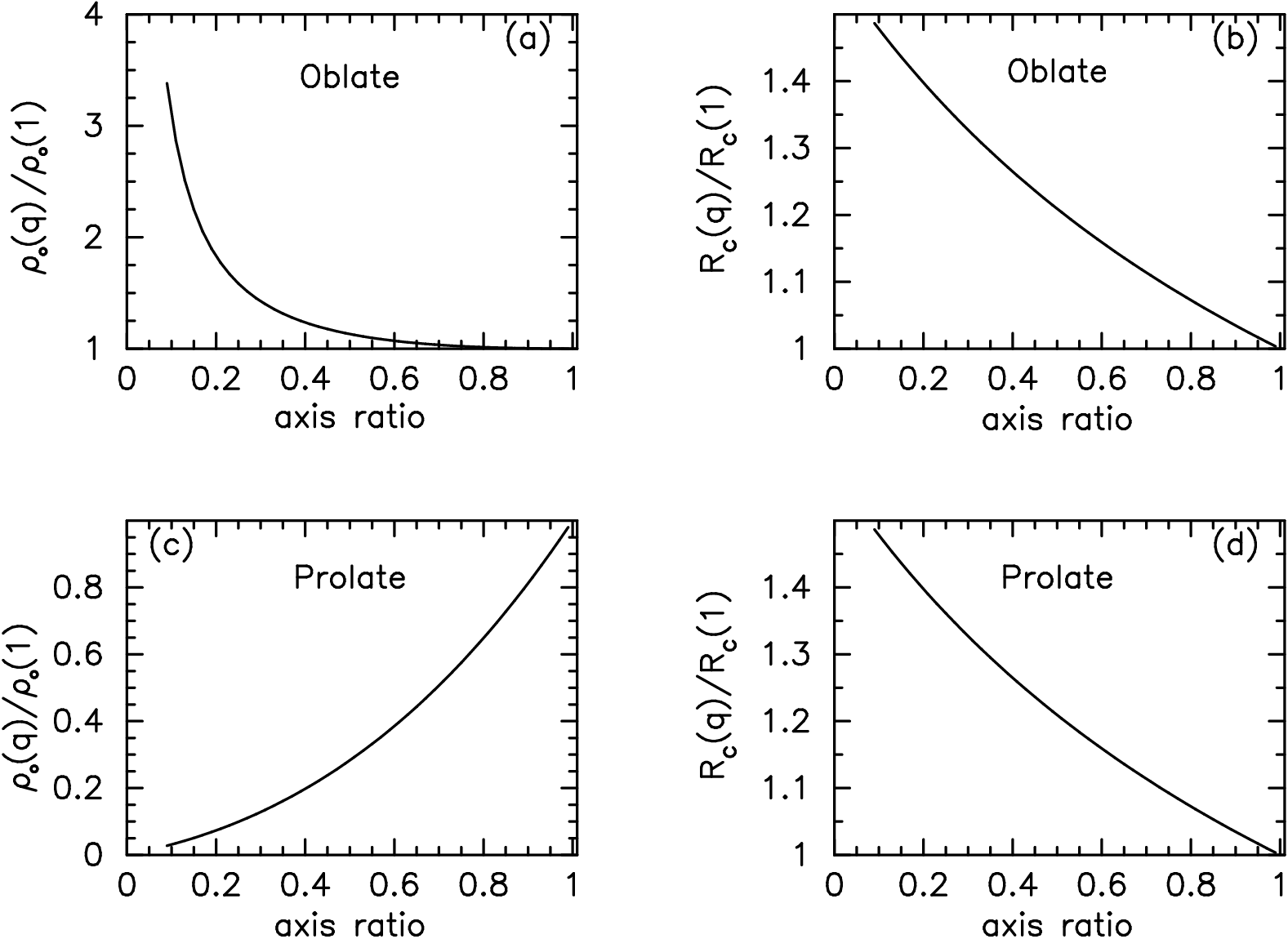}}}}
\bigskip

\noindent{\bf Figure 1.} This plot shows the variation of halo parameters as
 a function of its axis ratio (see eq.[7]). Figures 1(a) and (b) show the variation in the central density and the core radius, with oblateness (q = minor axis/major axis). Figures 1 (c) and (d) show the same as a function of prolateness (=minor axis/major axis). This dependence on axis ratio arises from keeping the mass within a thin spheroidal shell of the halo and the terminal velocity of the halo invariant of $q$.

\end{figure}

\begin{figure}
{\rotatebox{270}{\resizebox{8cm}{8cm}{\includegraphics{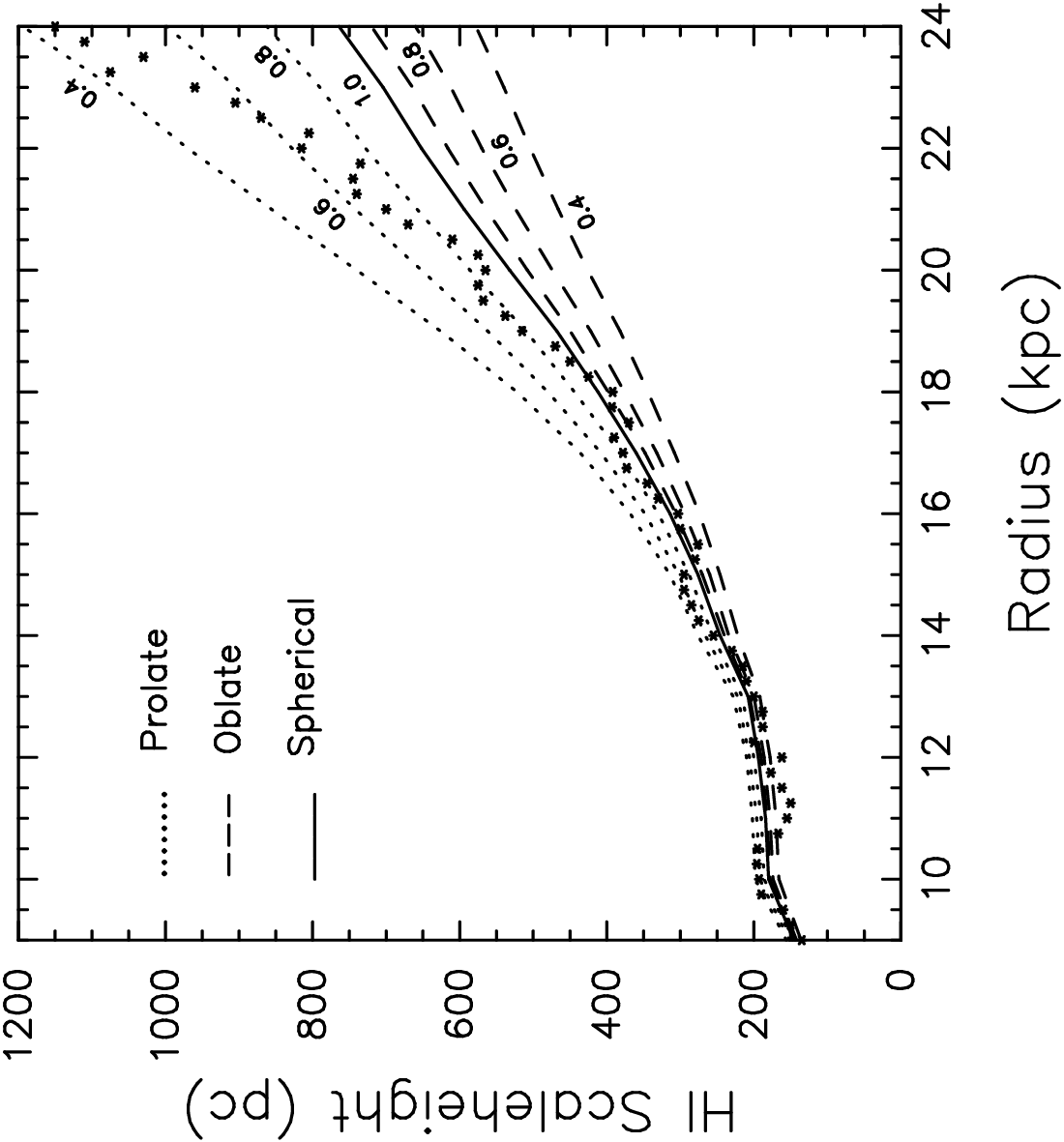}}}}
\noindent{\bf Figure 2.} The calculated HI scaleheight is shown as a function of the galactocentric radius in the outer region of the Galaxy. The shape of an initially spherical isothermal halo ($\rho_{\circ}$ = 0.035 M$_{\odot}$ pc$^{-3}$ ; R$_c$ = 5 kpc) is changed keeping its mass constant. This plot shows the results from our model for different axis ratios. The solid line is due to spherical shape ($q = c/a $= 1); the dashed lines are  due to oblate halos ($c/a$ = 0.8, 0.6, 0.4) and the dotted lines are due to prolate halos ($a/c$ = 0.8, 0.6, 0.4). The points show the observed values. Note that neither the oblate nor the prolate-shaped halos are clearly favoured by the data.
\end{figure}

\bigskip
\begin{figure}
{\rotatebox{270}{\resizebox{9cm}{9cm}{\includegraphics{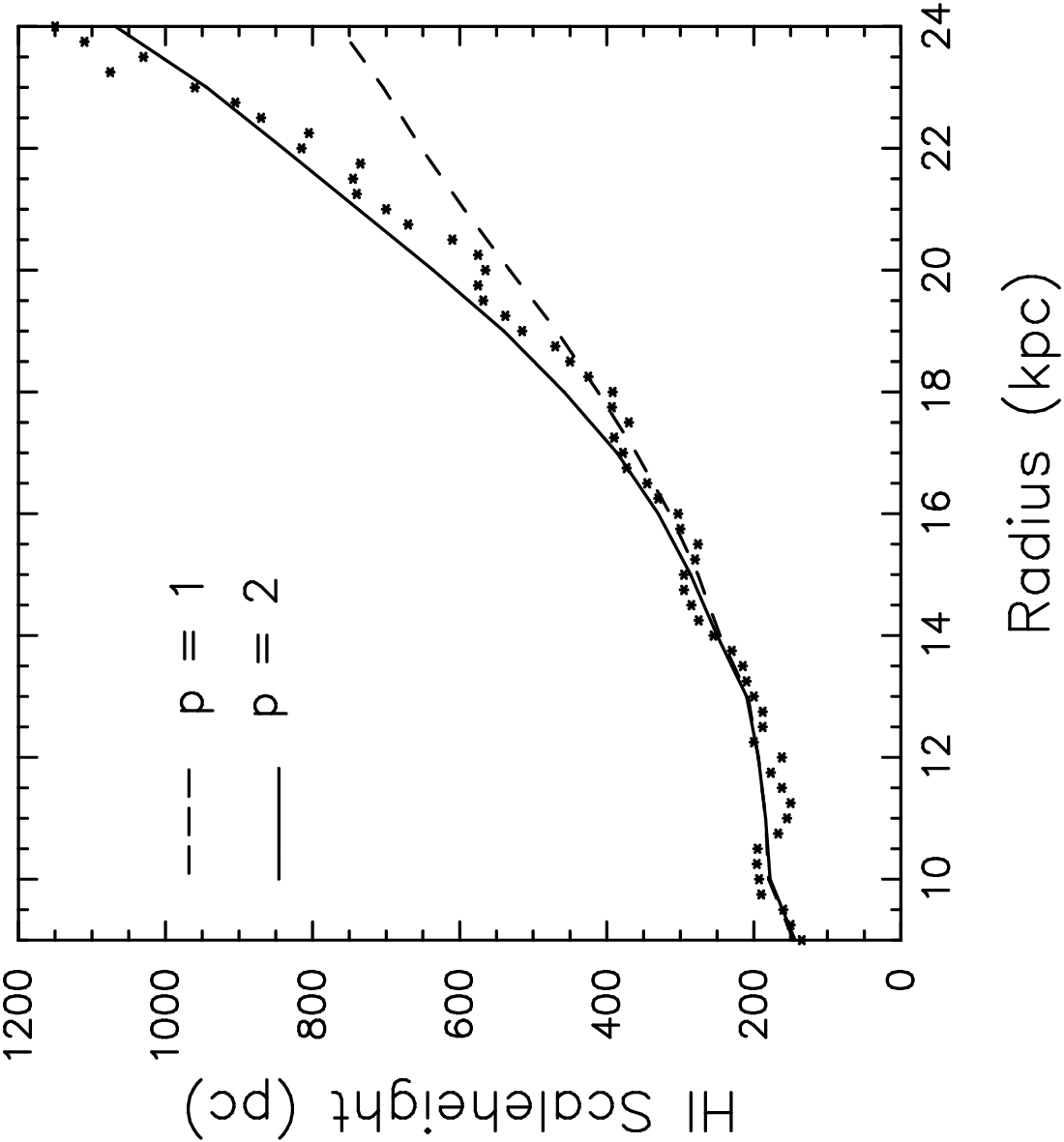}}}}
\bigskip

\noindent{\bf Figure 3.} The best fit (based on the least square value) for the observed HI scaleheight is given by a halo with $p$ = 2 where $p$ is the index in the halo density profile (solid line). For comparison, results for a typical isothermal halo ($p$ = 1) is also shown (dashed line). 

\end{figure}

\bigskip
\begin{figure}
{\rotatebox{270}{\resizebox{9cm}{9cm}{\includegraphics{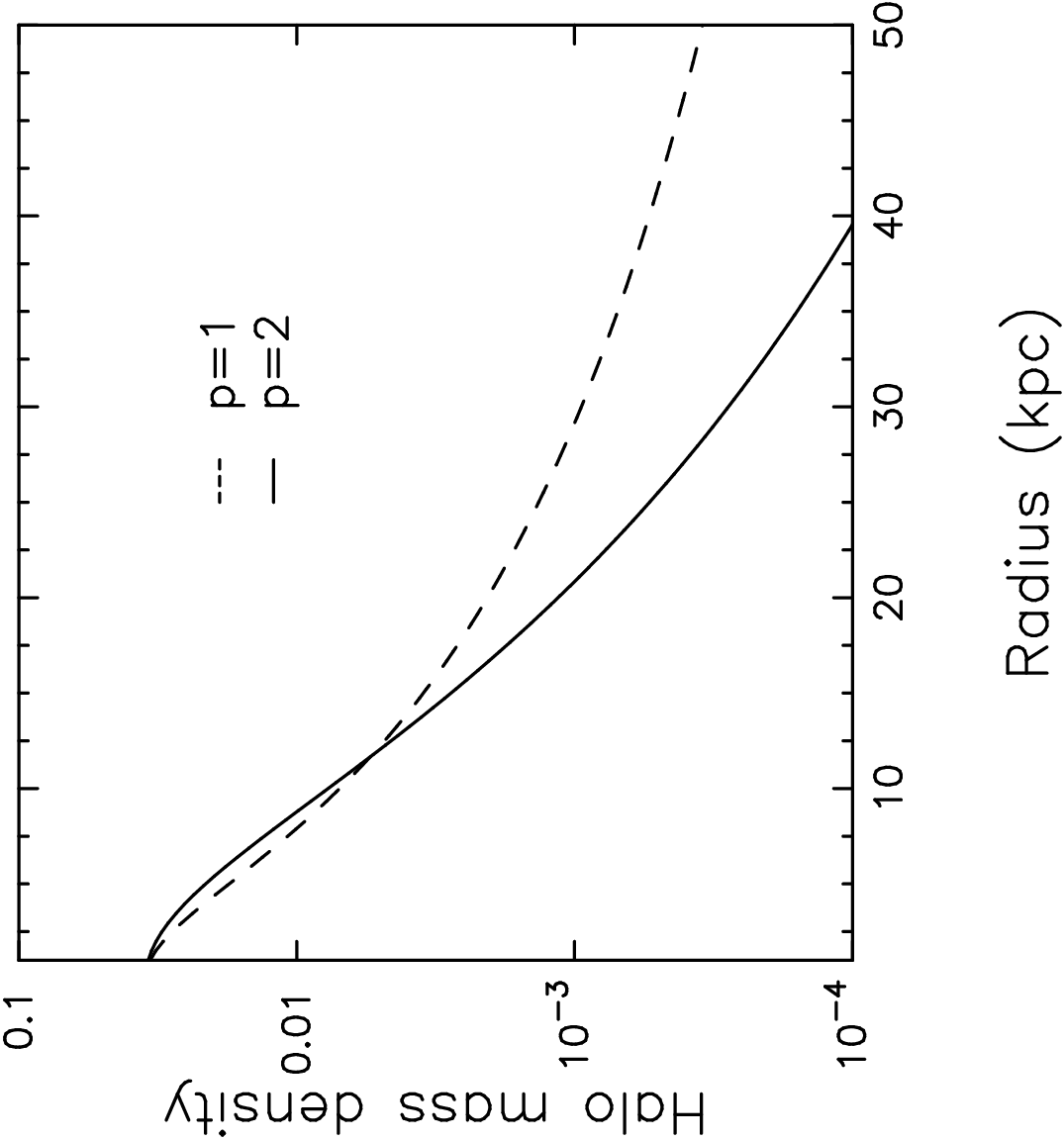}}}}

\noindent{\bf Figure 4a.} A log-normal plot of the halo mass density (in units of M$_{\odot}$pc$^{-3}$) as a function of the galactocentric radius, for the best-fit halo (p=2) and the isothermal halo (p=1). Although the two begin to differ around the optical edge of the stellar disk itself, the effect on HI scaleheight becomes noticeable only after 20 kpc (see Fig. 3). The corresponding p=1.5 profile (which is equivalent to the NFW density profile at large radii) falls between the two shown profiles but is closer to the p=2 profile.     

\end{figure}
\bigskip
\bigskip
\begin{figure}
{\rotatebox{270}{\resizebox{9cm}{9cm}{\includegraphics{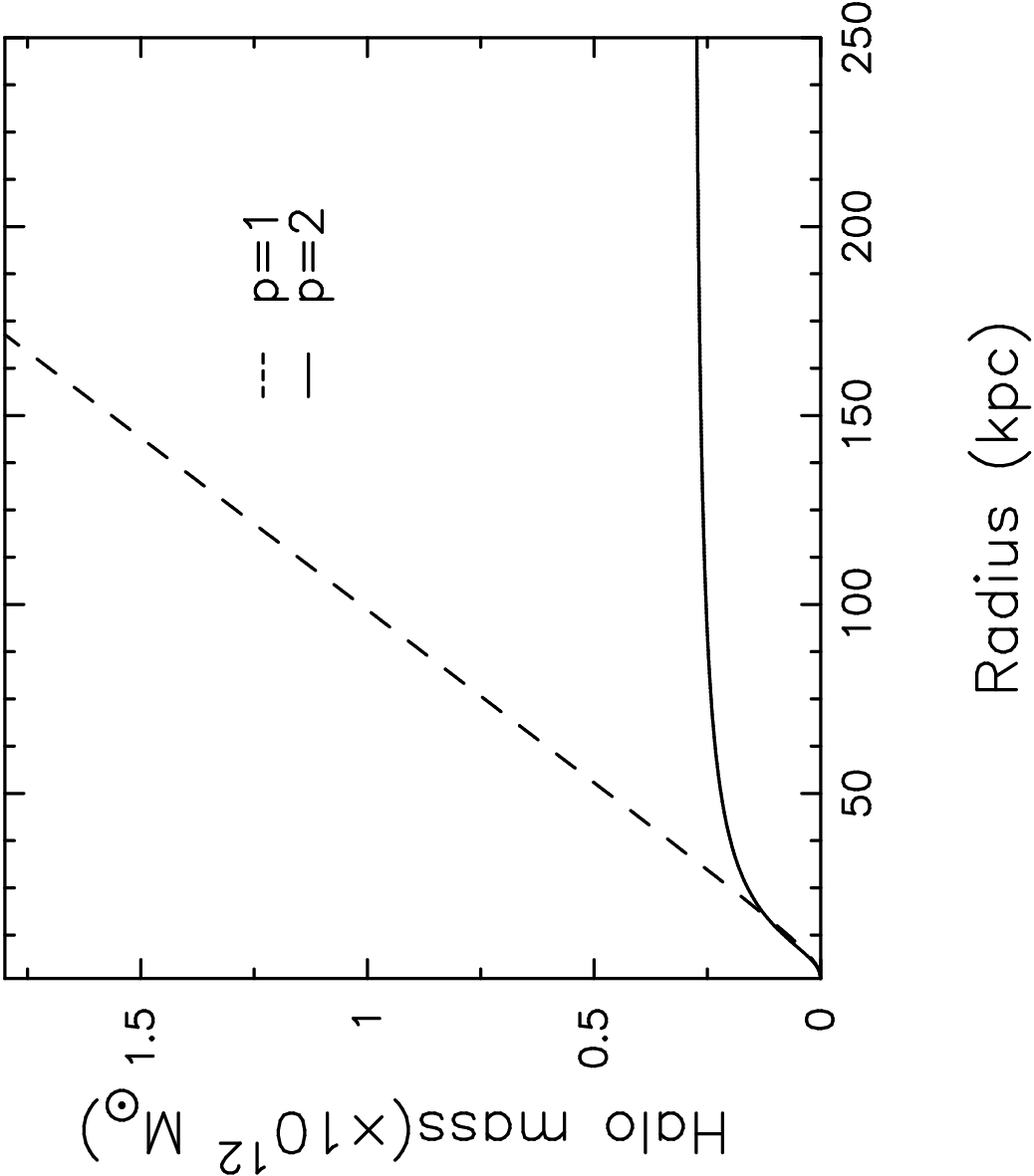}}}}

\noindent {\bf Figure 4b.} The behaviour of M(R) (mass contained within a spheroid of radius R) as a function of R, is shown here for the two kinds of halos. For an isothermal halo, 
the mass tends to infinity whereas for p=2, 
it tends converges to a finite value as R increases. 
The mass tends to infinity for p=1.5 as well, 
but rather gradually compared to that of p=1.

\end{figure}
\bigskip
\bigskip
\begin{figure}
{\rotatebox{270}{\resizebox{9cm}{9cm}{\includegraphics{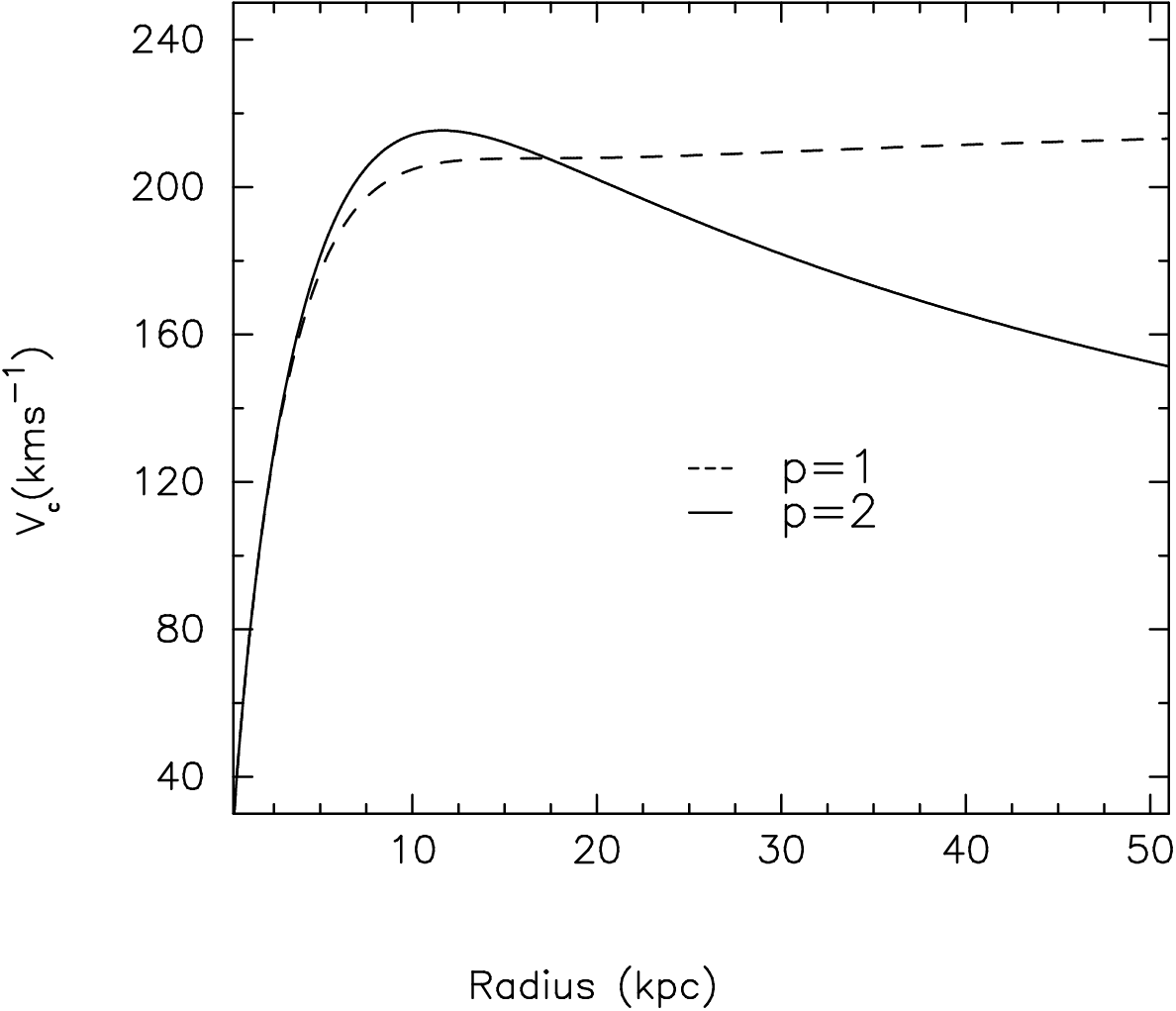}}}}

\noindent{\bf Figure 4c.} Rotation curves for the cases p=1
and p=2. For p=1, the rotation curve becomes flat at large radii
whereas for p=2, the curve begings to fall beyond a radius of
$\sim$ 12 kpc.

\end{figure}

\end{document}